# The uncollapsed $LaFe_2As_2$ phase: compensated, highly doped, electron-phonon coupled, iron-based superconductor

Ilaria Pallecchi [1], Akira Iyo [2], Hiraku Ogino [2], Marco Affronte [3], Marina Putti [4,1]

[1] CNR-SPIN, c/o Dipartimento di Fisica, via Dodecaneso 33, 16146 Genova, Italy
[2] National Institute of Advanced Industrial Science and Technology (AIST), 1-1-1 Umezono, Tsukuba, Ibaraki 305-8568, Japan
[3] Dipartimento di Scienze Fisiche, Informatiche e Matematiche, Università degli Studi di Modena e Reggio Emilia, and CNR-NANO, via G. Campi, 213/A, 41100 Modena, Italy
[4] Università di Genova, Dipartimento di Fisica, via Dodecaneso 33, 16146 Genova, Italy

**Abstract**
The recently discovered $LaFe_2As_2$ superconducting compound, member of the 122 family of iron pnictide superconductors, becomes superconducting below $T_c \approx 13K$, yet its nominal doping apparently places it in the extreme overdoped limit, where superconductivity should be suppressed. In this work, we investigate the normal state of magneto- and thermo-electric transport and specific heat of this compound. The experimental data are consistent with the presence of highly compensated electron and hole bands, with ~0.42 electrons per unit cell just above $T_c$, and high effective masses ~$3m_0$. The temperature dependence of transport properties strongly resembles that of conventional superconductors, pointing to a key role of electron-phonon coupling. From these evidences, $LaFe_2As_2$ can be regarded as the connecting compound between unconventional and conventional superconductors.

**1. Introduction**
Dome-shaped phase diagrams are a distinct hallmark of unconventional superconductivity. Iron-based superconductors are no exception to this rule. Specifically, in the so called 122 family, with general composition $AFe_2As_2$ (A=alkaline earth metal), superconductivity occurs in a doping range below ~0.2 electrons/Fe [1]. This assumption was apparently challenged when the new stoichiometric compound $LaFe_2As_2$ was synthesized and found to be superconducting below 12 K [2]. Indeed, this compound has a nominal doping of 0.5 electrons/Fe, which would place it in the dramatically overdoped regime, where superconductivity should be suppressed. Studying such odd-one-out may unveil the mysteries of unconventional superconductivity in iron pnictides.

Both $LaFe_2As_2$ and its isostructural $CaFe_2As_2$ derivative exist in two distinct crystallographic phases, namely the "collapsed" phase with shorter c-axis and the "uncollapsed" phase with elongated c-axis, and bulk superconductivity appears around ~12 K in just one of these two phases [2,3,4]. Indeed, the structural instability between collapsed and uncollapsed phases is typical of $AFe_2As_2$ compounds (A = alkali or alkaline-earth metal), where the drastic changes in lattice parameters are accompanied by significant changes of the electronic properties, related to the underlying changes of the dimensionality of the electronic band structure and changes of the Fe magnetic moment [5,6]. In the case of $LaFe_2As_2$, the as-synthesized "collapsed" compound is not superconducting, while the 500°C annealed "uncollapsed" compound is, yet none of the two phases exhibits long-range magnetic ordering [2]. The phase diagram of the uncollapsed system was explored in the chemically substituted $(La_{0.5-x}Na_{0.5+x})Fe_2As_2$ compound [7,8]. Here, the formal valence of FeAs layer is controlled linearly by the single parameter x from hole-doping (x>0) to electron-doping (x<0) region, with x=0 corresponding to the Fe formal valence +2, typical of Fe-pnictides parent compounds. Indeed, stripe-type antiferromagnetic order below $T_N$ = 130 K was found for x=0, while superconductivity of multigap nature was found below $T_c$~9.4 K and $T_c$~27 K for x=-0.5 and x=+0.3, respectively. On the other hand, no intrinsic bulk superconductivity was detected in $La_{0.4}Na_{0.6}Fe_2As_2$ by either Co doping or application of pressure, even if in both cases suppression of the antiferromagnetic ordering was obtained [9].

To shed light on this system, structural, magnetic, and electronic properties of $LaFe_2As_2$ were calculated *ab initio* by Mazin et al. [10], indicating that the uncollapsed phase carries a strong short-range magnetism, which ultimately drives the superconducting transition. The puzzle related to the absence of any hole pockets near the zone center [2,7], which is virtually ubiquitous in superconducting iron pnictides, was solved by calculations that showed that the orbitals relevant for the low-energy physics are not the usual $d_{xz}$ and $d_{yz}$, which are almost completely filled, but rather the $d_{xy}$ and $d_z$ ones [10]. Indeed, $d_{xy}$ orbital forms a quasi-two dimensional cylinder at the zone center, much resembling the hole pocket in other iron pnictides [10,11]. Experimental evidence of such hole pockets is argued [2,7], yet still lacking so far. As for the puzzle of dramatic overdoping, in the calculated band structure, the $5d_{xy}$ orbital is strongly hybridized with the As 4p orbital and has a huge dispersion [10] and both these factors cause an absorption of electrons by other bands well below the Fermi level, determining a reduced effective doping, as compared to the nominal 0.5 electrons/Fe. Even if the exact number of carrier could not be estimated precisely, due to strong hybridization, upper and lower limits of 0.47 and 0.22 electrons/Fe were given in ref. [10], which certainly places the compound in the overdoped regime, but well below 0.5 electrons/Fe.

Further insight into the plausible pairing mechanisms of $LaFe_2As_2$ were given from *ab initio* calculations of ref. [12], where the proximity of the narrow Fe $d_{xy}$ band to the Fermi level was identified as a key feature for the appearance of superconductivity in the uncollapsed phase of $LaFe_2As_2$, with respect to the collapsed phase. In this work, it was suggested that correlation and enhanced scattering in the $d_{xy}$ band result in intense low energy spin fluctuations, that provide glue for unconventional Cooper pair formation.

In this scenario, an experimental input is necessary to find a place for superconducting $LaFe_2As_2$ in a doping-$T_c$ phase diagram and possibly reconcile its description with that of other iron pnictides.

In this work, we measure normal state transport properties and specific heat in polycrystalline uncollapsed $LaFe_2As_2$ samples, and find evidence of electron and hole bands with high effective masses contributing to transport. We extract electronic parameters in a two-band framework and find that this compound is highly overdoped and highly compensated. While the compensated character makes $LaFe_2As_2$ akin to other iron-based superconductors, such as isovalent substituted chalcogenides [13,14], the highly overdoped character and the distinctive temperature dependence of resistivity and Seebeck coefficient typical of phonon-mediated superconductors make it the odd-one-out among iron-based superconductors.

**2. Magnetoelectric and thermoelectric transport properties**

Uncollapsed $LaFe_2As_2$ samples were synthesized using a high-pressure and high-temperature synthesis method with subsequent annealing as described in the ref. [2]. Magnetotransport and Seebeck measurements were carried out in a Physical Property Measurement System (PPMS) by Quantum Design, in applied magnetic fields up to 9T and at temperatures down to 5 K. Two samples were fully characterized. Since very similar behavior was observed in the two samples, in this paper we present data on one of them.

*(a) Resistivity*

In Fig. 1, the metallic resistivity of $LaFe_2As_2$ is shown. The room temperature resistivity is 80 $\mu\Omega$ cm and the residual resistivity is of the order 1 $\mu\Omega$ cm, so that a residual resistivity ratio ~80 can be evaluated. Notably, the resistivity does not flatten in the temperature range just above $T_c$, indicating that scattering by phonons is the dominant mechanism, even at the lowest temperatures.

The onset of superconductivity is at $T_c$=13.1 K and the transition width is ~1 K. In the upper left inset, the resistive transition at different applied fields is shown. With applied field, the transition onset shifts monotonically to lower temperatures and the transition width increases monotonically, as it is visible in the plot of upper critical field $H_{c2}$ and irreversibility field $H_{irr}$, in the lower right inset of Fig. 1. The $H_{c2}$ and $H_{irr}$ slopes are -1.82 T/K and -1.38 T/K, in agreement with ref. [2]. The upper inset in Fig. 1, presenting zero-field and in-field transitions, also shows a well visible departure of normal state resistivity of the zero-field curve from those in-field. On the other hand, for all the curves measured in fields equal or larger than 0.25 T the resistivity changes very little with applied field. This behavior was observed in both our measured samples.

As shown in the main panel of Fig. 1, the resistivity curve in the normal state is described by a generalized Bloch-Grüneisen law, typical of metals [15,16]:

$$\rho(T) = \rho_0 + \rho_{ph}(T) \quad \text{with} \quad \rho_{ph}(T) = (m-1)\rho'\Theta_R \left(\frac{T}{\Theta_R}\right)^m \int_0^{\Theta_R/T} \frac{z^m}{(1-e^{-z})(e^z-1)} dz \qquad (1)$$

In the low temperature range T < 35 K, the experimental resistivity curve follows a power law $\rho(T) = \rho_0+\text{const} \times T^m$ with $\rho_0$~1 $\mu\Omega$ and m~3. A similar behavior has been observed in $MgB_2$ and it is typical of multiband systems [17]. By fixing m=3, the temperature range up to ~220 K can be fitted with eq. (1), while at higher temperatures, the experimental curve bends with respect to the Bloch-Grüneisen law. This tendency to saturation has been observed in A15 superconductors [18,19] and it is typical of metals with large electron-phonon coupling; indeed when the resistivity rises steeply with increasing temperature, the mean free path decreases and approaches the lattice spacing, which sets a limit for further increase of resistivity, according to the Ioffe-Regel criterion [20]. We estimate that the mean free path at 300K is ~1 nm, comparable to the lattice parameters [2].

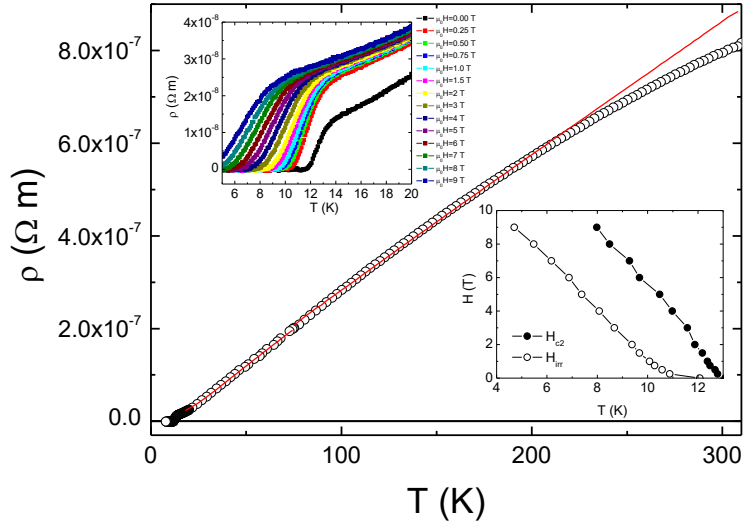

**Figure 1:** Resistivity of $LaFe_2As_2$. The continuous red line represents the fit with the generalized Bloch-Grüneisen law eq. (1). Upper left inset: resistivity curves in different magnetic fields up to 9 T. Lower right inset: $H_{c2}$ and $H_{irr}$ extracted from the resistive transitions with the criteria of 90% and 10% of the normal state resistivity, respectively.

*(b) Magnetoresistance and Hall effect*

Fig. 2 displays isothermal magnetoresistivity $(\Delta\rho(H)-\rho(H=0))/\rho(H=0)$, measured at different temperatures from 25 K to 300 K. At all the temperatures, magnetoresistivity is positive. For temperatures $\geq$ 50 K, $\Delta\rho$ decreases in magnitude with decreasing temperature, being around 1% at 9 T and 50 K and 0.25% at 9 T and room temperature. These curves can be described by the semiclassical model of magnetoresistivity proportional to the $B^2$. The curve at 25 K, magnified in the low field range in the inset of Fig 2, exhibits a different behavior which cannot be described in the semiclassical cyclotron framework, and is rather reminiscent of a weak antilocalization mechanism, with a sharp dip at low field and saturation at higher fields. The magnitude of $\Delta\rho$ is larger than 10%, for fields above 0.25 T, which is reconciled with the behavior of resistive transitions presented above. This behavior disappears with increasing temperature, so that above 50 K only the cyclotronic mechanisms survives.

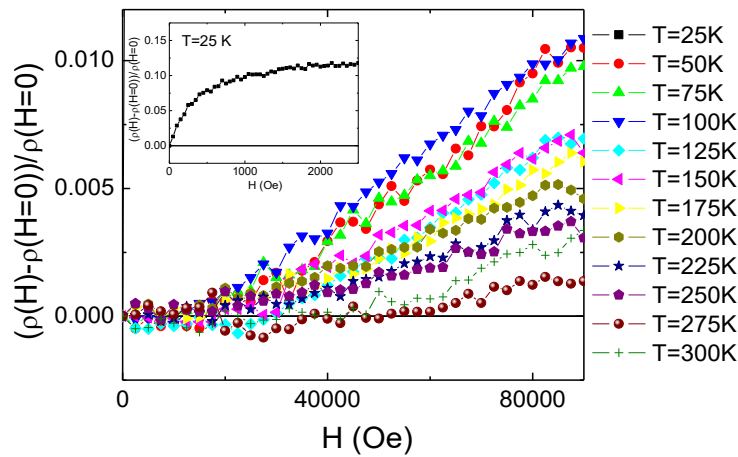

**Figure 2:** Isothermal magnetoresistivity of LaFe$_2$As$_2$ at different temperatures. The curve at 25 K is shown in the inset magnifying the low field regime.

The Hall resistance curves measured at different temperatures are presented in Fig. 3. The slope is negative, indicating that the dominant charge carriers are electrons. However, the most noteworthy feature is the well visible non-linearity, at all the temperatures except for 25 K. Such non-linearity is an unambiguous evidence of the presence of a hole band at the Fermi level, that participates in transport, as it is the case of most of the other 122 superconducting compounds.

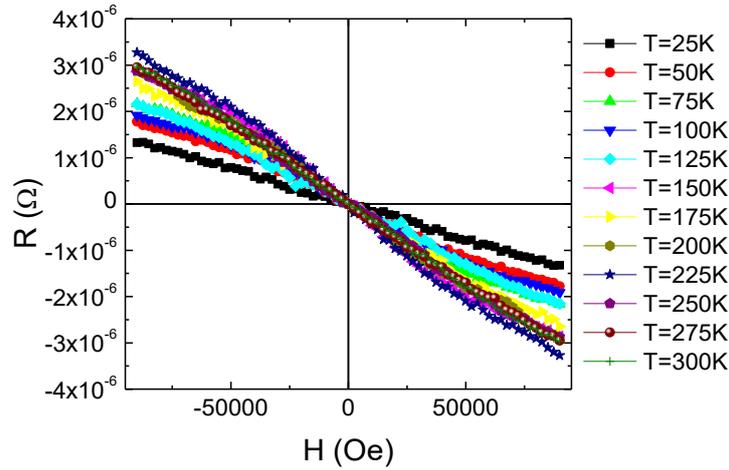

**Figure 3:** Hall resistance of LaFe$_2$As$_2$ at different temperatures.

*(c) Seebeck Effect*

In Fig. 4, the temperature dependence of the Seebeck coefficient S of LaFe$_2$As$_2$ is presented. The negative sign of S is consistent with the negative sign of the Hall effect. The monotonic temperature dependence of S and its small value ~27.5 μV/K at 300K are consistent with the metallic character of this compounds. More specifically, as for most of clean metals, S is composed of a linear diffusive contribution plus a broad bump at low temperatures, centered around 50-60 K in Fig. 4, which can be ascribed to a phonon drag contribution. For temperatures much smaller than the Debye temperature $\Theta_D$ the phonon drag Seebeck coefficient is expected to be proportional to the phonon contribution to the specific heat [21], thus behaving as ~$T^3$. We then assume that the experimental S is a sum of the diffusive Seebeck $\propto T$ plus the phonon drag Seebeck $\propto T^3$, as previously done for the conventional superconductor MgB$_2$ [15]:

$$S = AT + BT^3 \qquad (2)$$

Plotting S/T versus $T^2$, as in the inset of Fig. 4, we can identify a linear regime in the 40 K-85 K temperature range and extract the diffusive and phonon drag coefficients A and B, as the intercept and the slope, respectively. The obtained values are A≈-0.186 μV/K$^2$ and B≈-5.13x10$^{-6}$ μV/K$^4$. Considering now the B coefficient which represents the amplitude of the phonon drag term, we note that the ratio B/A≈3 x10$^{-5}$ K$^{-2}$ is of the same order of c MgB$_2$ (B/A≈7 x10$^{-5}$ K$^{-2}$) [15]. This observation of a Seebeck term proportional to $T^3$ with a reliable order of magnitude is a clear evidence that phonon drag contributes significantly to Seebeck effect; this generally occurs in pure metals with strong electron-phonon coupling, where phonon scattering is the dominant interaction of electrons, as for the case of MgB$_2$.

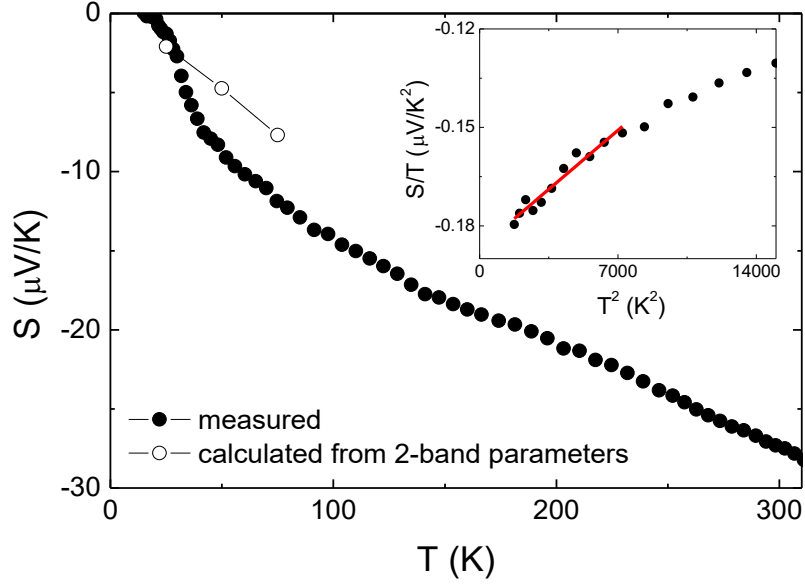

**Figure 4:** Measured Seebeck coefficient of $LaFe_2As_2$ (filled symbols). The diffusive Seebeck coefficient calculated from the two-band parameters is also shown in the main panel (open symbols). Inset: S/T versus $T^2$ plot, with a linear regime identified in correspondence of the temperature range 40-85K, as described by eq. (2), with fitting parameters A≈- 0.186 µV/K² and B≈5.13x10⁻⁶ µV/K⁴.

### 3. Specific heat measurements

Heat capacity measurements were performed with a PPMS system of the Quantum Design on a polycrystalline sample of mass 71.8mg by the relaxation method with typical temperature pulse of $\Delta T \approx 1\%$ with respect to the bath temperature.

In Fig. 5, the temperature dependence of the specific heat $c$ is presented. The lower right inset shows that the $c$ value, normalized to the universal gas constant R, tends to saturate at high temperature at a value slightly larger than 15 R, as expected from the Dulong–Petit law for the lattice contribution 3 x $N_{at}$ x R where $N_{at}$=5 is the number of atoms per unit cell. The excess to this value obviously comes from the electronic and magnetic contributions.

The molar specific heat $c$ measured in zero field and in a 7 T field below 20 K, is plotted as c/T versus the squared temperature $T^2$ in the main panel of Fig. 5. Below $T_c$, the zero field data exhibit an onset of a broad bump, related to the superconducting transition. For the 7 Tesla data, the transition is broadened is shifted to lower temperatures. These zero-field and in-field superconducting transitions are better evidenced in the plot of the electronic superconducting contribution to the specific heat $c_{es}/T=c/T-(\gamma+\beta T^2)$ versus T, shown in the upper left inset of Fig. 5. It is seen that the onsets of the transitions are consistent with the $H_{c2}$ data extracted from the superconducting onsets of the resistivity curves (lower right inset of Fig. 1), indicated by arrows at $T_c \approx 13K$ in zero field and at $T_c \approx 9.5K$ in 7T field.

Between 10 K and 20 K, the in-field c(7T) data in the normal state fits well the law:
$$c(T) = \gamma T + \beta T^3 \qquad (3)$$
where $\gamma$ is the electronic specific heat, also known as Sommerfeld coefficient, and the $c/T \propto T^2$ term is the Debye contribution of acoustic phonons. The best fit of 7 Tesla data between 10 K and 20K gives $\gamma \approx 35$ mJ mol⁻¹ K⁻² and $\beta \approx 0.73$ mJ mol⁻¹ K⁻⁴, with a 5% uncertainty on both coefficients $\gamma$ and $\beta$ related to slight variations of the fitting range. The molar electronic specific heat $\gamma \approx 35$ mJ mol⁻¹ K⁻² is quite large a value for a good metal, indicative of a high effective mass. The slope $\beta$=0.73 mJ mol⁻¹ K⁻⁴ is related to the phonon spectrum and, in the low temperature limit where only acoustic phonons contribute to the specific heat, it is related to the Debye temperature $\Theta_D$ by the relation:

$$\Theta_D = \sqrt[3]{\left(\tfrac{12}{5}\pi^4 R\right)/\beta} \qquad (4)$$

From eq. (4), we obtain for the Debye temperature $\Theta_D \approx 140$ K.

We finally note that in the upper left inset of Fig. 5 the magnitude of $c_{es}/T$ maximum is around 20 mJ mol$^{-1}$ K$^{-2}$. Considering that for a homogeneous BCS superconductor, the difference between superconducting and normal electron contributions to specific heat at $T_c$ normalized to $\gamma T_c$ is expected to be $\frac{C_{es}(T_c)}{\gamma T_c} \approx 1.43$, our values point to a bulk superconductor, although not fully homogeneous.

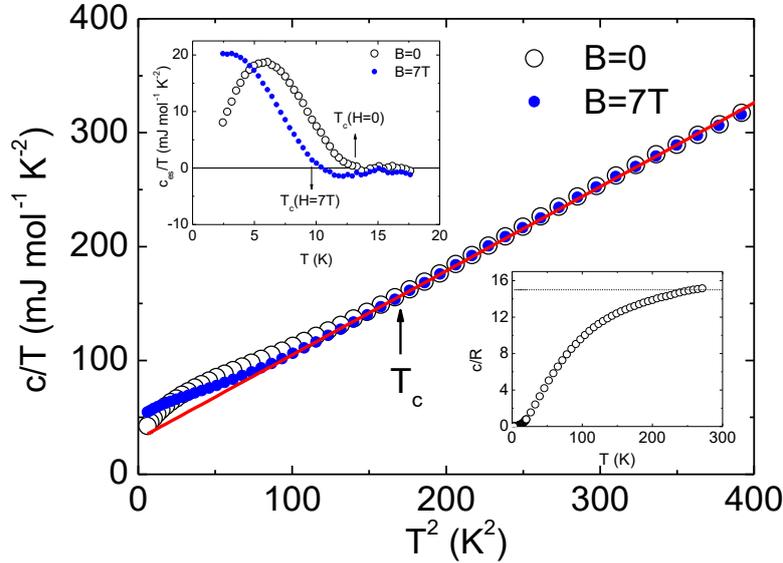

**Figure 5:** Specific heat $c_p$ of LaFe$_2$As$_2$ in zero field and in a 7 T field, plotted as $c/T$ versus $T^2$. The continuous red line represents a linear fit to the 7 Tesla data in the temperature range between $T_c$ and 20K, with best fit coefficients $c/T$ [mJ mol$^{-1}$ K$^{-2}$] = (35 + 0.73 T$^2$) [mJ mol$^{-1}$ K$^{-2}$]. In the upper left inset, the electronic superconducting contribution to the specific heat $c_{es}/T=c/T-(\gamma+\beta T^2)$, which better evidences the superconducting transition, is plotted versus T. In the lower right inset, c in units of R is plotted versus T and the horizontal dashed line represents the value 3 x N$_{at}$ x R=15 R, expected from the Dulong–Petit law for the lattice contribution.

## 4. Data analysis and discussion

*(a) Carrier densities and mobilities*

Normal state magnetotransport is revealing of band parameters if analyzed in the proper framework, thus offering the opportunity of clarifying the real nature of this apparently peculiar member of the 122 iron based superconductors. We first analyze data in a simple single band framework. In the upper panel of Fig. 6, carrier density and mobility are extracted respectively from the linear fit of Hall resistance curves and from the inverse product of carrier density and resistivity. The carrier density varies from 2x10$^{22}$ cm$^{-3}$ at 25 K to 9x10$^{21}$ cm$^{-3}$ at 300 K. Temperature dependence of carrier density is not expected in a single band framework. In addition, the absolute values are unrealistically large, the 25 K values pointing to 3.5 electrons per unit cell. These two observations, together with the non-linearity Hall resistance curves (see Fig. 3), indicate that the single band description is inadequate, thereby a two-band analysis, with one electron band and one hole band, must be carried out, instead. In order to determine carrier densities and mobilities of the two bands, $n_e$, $n_h$, $\mu_e$ and $\mu_h$ at each fixed temperature, we need 4 equations. The expression of the non-linear Hall resistance is written as:

$$R_{xy} = \frac{Vol}{q} \frac{(-\mu_e^2 n_e + \mu_h^2 n_h) + (-n_e + n_h)(\mu_e \mu_h B)^2}{(\mu_e n_e + \mu_h n_h)^2 + (-n_e + n_h)^2 (\mu_e \mu_h B)^2} B \qquad (5)$$

where the sign of the charge carriers is already made explicit, so that the parameters $n_e$, $n_h$, $\mu_e$ and $\mu_h$ are all positive. In eq. (5), Vol is the unit cell volume, q is the positive electronic charge and B is the applied field in Tesla. Despite fitting the experimental curves of Fig. 3 with eq. (5) could in principle provide 3 coefficients and thus 3 equations, our experimental curves can be well fitted by just two parameters:

$$R_{fit} = \frac{\alpha B + \xi B^3}{1+\delta B^2} \approx \alpha\mu_0 B + (\xi - \alpha\delta)B^3 \quad (6)$$

where the tentative assumption $\delta B^2 \ll 1$ is done. We obtain the other two equations from the values of resistivity and of cyclotron magnetoresistivity at 9T, expressed respectively as:

$$\rho = \frac{Vol}{q}\frac{1}{\mu_e n_e + \mu_h n_h} \quad (7)$$

$$\frac{\rho(B)-\rho(0)}{\rho(0)} \approx \frac{n_e n_h \mu_e \mu_h (\mu_e + \mu_h)^2}{(\mu_e n_e + \mu_h n_h)^2} B^2 \quad (8)$$

Note that eq. (8) is truncated at the leading order in $B^2$, as applicable to the low magnetorsistivies in Fig. 2. Note also that in eq. (8) contribution of bands of different signs (holes and electrons) are additive, while contributions of bands of the same sign would subtract as $\propto (\mu_1 - \mu_2)^2$.

Hence by combining the equations for the linear and cubic fitting coefficients of $R_{xy}$ given by eqs. (5) and (6), for $\rho$ given by eq. (7) and for $\frac{\rho(B=9T)-\rho(0)}{\rho(0)}$ given by eq. (8), we obtain $n_e$, $n_h$, $\mu_e$ and $\mu_h$ at each temperature, as displayed in the bottom panel of Fig. 6. The datum at 25 K is missing because we have neither the cyclotron magnetoresistance (see inset of Fig. 2) nor the non-linear Hall resistance coefficient ($\xi-\alpha\delta$) in eq. (6), as the Hall resistance at 25 K is linear in the field (see Fig. 3). Note that the solution is not univocal in principle, because the experimental curvature of the flex point in the $R_{xy}$ curve can be reproduced with infinite choices of the parameters $\xi$ and $\delta$, however solutions are only found for $\delta \ll \xi/\alpha$ (and $\delta B^2 \ll 1$, as assumed *a priori*), that is for $R_{fit} \approx \alpha\mu_0 B + \xi B^3$ in eq. (6). Note also that by dropping the assumption $\delta B^2 \ll 1$, and assuming instead $\xi \ll a/B^2$, that is using the fitting equation $R_{fit} = \frac{\alpha B + \xi B^3}{1+\delta B^2} \approx \frac{\alpha B}{1+\delta B^2}$, no solution is found for $n_e$, $n_h$, $\mu_e$ and $\mu_h$. We also tried to analyze data by assuming two electron bands, but no solution was found in this case either, thus confirming the existence of a hole band contribution to transport.

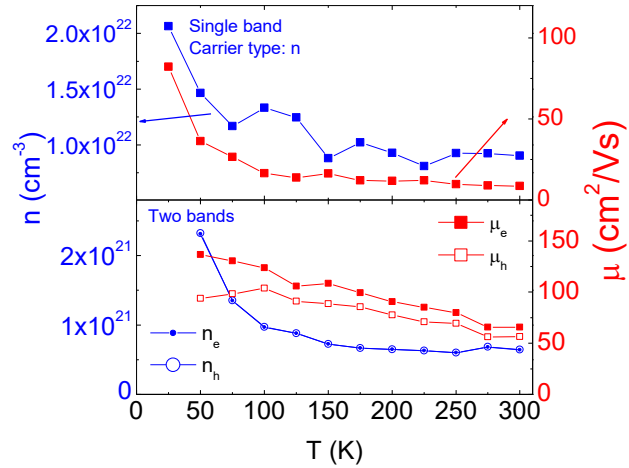

**Figure 6:** Carrier concentrations and mobilities extracted from data analysis in a single band framework (upper panel) and in a two-band framework (lower panel).

From these results we conclude that : *i)* LaFe$_2$As$_2$ is an highly compensated compound, with carrier densities of the hole and electron bands that differ by around 1%; *ii)* the carrier densities of the hole and electron bands are ~0.42 electrons per unit cell at 50 K and ~0.12 electrons per unit cell at room temperature, hence in the highly overdoped regime at low temperature T=50 K; *iii)* electron mobility is larger than hole one by a factor 1.5 at low temperature and 1.2 at room temperature, with both $\mu_e$ and $\mu_h$ decreasing by a factor ~2 in this temperature range. The compensation up to ~99% of hole and electron bands needs to be further discussed. Although charge compensation is not unusual in iron based superconductors [13,14], compensation up to ~99% is quite remarkable. On the other hand, the extremely low experimental values of the Hall effect, measured in both our samples, can only be explained only by either unrealistically large carrier densities, as in the single band analysis, or by very high carrier compensation of bands of different signs. The latter seems to be the case, as indicated by our analysis with no free parameters. Such high degree of compensation should yield non-saturating quadratic magnetoresistance up to very high fields ≥ 35 T, which should be checked in future experiments.

*(b) Effective masses*
The effective mass can be extracted from the Sommerfeld coefficient γ, which can be expressed as the sum of electron and hole contributions $\gamma_e$ and $\gamma_h$ as:

$$\gamma = \gamma_e + \gamma_h = \frac{\pi^2 K_B^2 N_{av}}{2}\left(\frac{1}{E_{F(e)}} + \frac{1}{E_{F(h)}}\right) \quad (9)$$

where $N_{av}$ is the Avogadro number and $E_F$ the Fermi energy, which for three-dimensional parabolic bands can be expressed as $E_F = \frac{\hbar^2}{2m_{eff}}(3\pi^2 n)^{\frac{2}{3}}$. From the two-band analysis it turns out that electron and hole mobilities are nearly equal, therefore it is reasonable to assume nearly equal effective masses $m_{eff(e)} \approx m_{eff(h)}$, which implies nearly equal Fermi energies, given that $n_e \approx n_h$. Hence the value γ = (35±2) mJ mol$^{-1}$ K$^{-2}$ found from specific heat is reproduced by assuming effective masses $m_{eff(e)} \approx m_{eff(h)} \approx 3\, m_0$. This $m_{eff}$ value is similar to what is on average obtained from de Haas–van Alphen [22] and Angle resolved Photoemission Spectroscopy (ARPES) [23] measurements on 122 superconductors.

The analysis of the Seebeck effect provides additional information. The diffusive contribution to the Seebeck coefficient can be calculated in the two-band framework. The conductivities $\sigma_e$ and $\sigma_h$ weigh the sum of band contributions $S_e$ and $S_h$ to the Seebeck coefficient, according to the expression:

$$S = \frac{\sigma_e S_e + \sigma_h S_h}{\sigma_e + \sigma_h} \quad (10)$$

Here the diffusive contributions of electron and hole bands to the Seebeck coefficient can be expressed by the Mott formula for metals, predicting a linear temperature dependence:

$$S_d = \pm \frac{\pi^2 K_B^2 T}{3q}\frac{\sigma'}{\sigma} \quad (11)$$

where $K_B$ is the Boltzmann constant, the sign +/- applies to holes/electrons, and σ and σ' are the conductivity and its energy derivative calculated at the Fermi level $E_F$. By expressing σ and σ' with the Drude formula, the ratio σ'/σ results in the sum of logarithmic derivatives of carrier density ∂ln(n)/∂E and of scattering time ∂ln(τ)/∂E. Given the energy dependence of the density of states N(E)∝$E^{1/2}$, we get ∂ln(n)/∂E=3/2. The energy dependence of τ can be modelled by a power law $\tau^\eta$, where the exponent η depends on the scattering mechanism and it is η=-1/2 for scattering by acoustic phonons (∂ln(τ)/∂E=-1/2) and η=0 for scattering by impurities. With these assumptions, eq. (11) is written as:

$$S_d = \pm\left(\frac{3}{2} + \eta\right)\frac{\pi^2}{3}\frac{K_B^2}{q}T\frac{1}{E_F} \quad (12)$$

Finally introducing the Fermi energy expression, the two-band diffusive Seebeck coefficient in eq. (10) can be written in terms of the band parameters $n_e$, $n_h$, $\mu_e$, $\mu_h$ and $m_{eff(e)} \approx m_{eff(h)}$ as:

$$S_d = C * T * \frac{\left(n_e^{1/3}\mu_e \frac{m_{eff(e)}}{m_0}\right) - \left(n_h^{1/3}\mu_h \frac{m_{eff(h)}}{m_0}\right)}{n_e\mu_e + n_h\mu_h} \quad (13)$$

where $C = \left(\frac{3}{2}+\eta\right)\frac{\pi^2}{3}\frac{K_B^2}{q}\frac{2m_0}{\hbar^2(3\pi^2)^{\frac{2}{3}}}$ is a dimensional constant and $m_0$ is the bare electron mass.

Using the two-band parameters extracted from the fit of specific heat and magnetotransport data, the diffusive Seebeck coefficient $S_d$ can be calculated by eq. (13). This calculated $S_d$ is plotted as open symbols in Fig. 4, assuming η=0 in the low temperature limit of scattering by impurities. Notably, given the strong compensation between the electron and hole contribution which suppresses $S_d$, the consistency between the measured S and $S_d$, can be obtained only with large effective masses, at least ~3$m_0$, hence the analysis of the Seebeck curve is an independent and consistent evaluation of $m_{eff}$.

We can now compare the slope of the linear fit $dS_d/dT \approx -0.11$ µV/K$^2$ of diffusive Seebeck calculated by eq. (13) with the intercept A≈-0.19 µV/K$^2$ evaluated by eq. (2) form the temperature dependence of the Seebeck effect. The magnitude of A is larger than $dS_d/dT$, however this is pretty plausible, considering that the linear temperature dependence predicted by the Mott law is itself an oversimplification, which does not describe experimental data when different scattering mechanisms come into play across the temperature range [24] and possible renormalization effects [15,25].

*(c) Electron-phonon coupling constant*

So far, we collected several evidences of the prominent role of electron-phonon coupling in this LaFe$_2$As$_2$ making it mandatory to give un estimation of the electron-phonon coupling from transport properties. An approximate evaluation of the transport electron-phonon coupling $\lambda_{tr}$ can be obtained by the resistivity coefficient $\rho'$ in eq. (1), which can be expressed as [15]:

$$\rho' = \frac{m_{eff}}{nq^2} \frac{2\pi K_B}{\hbar} \lambda_{tr} \tag{14}$$

Therefore $\lambda_{tr}$ can be evaluated using the values of carrier density and effective mass as in the above analysis and the coefficient $\rho'$ extracted by the Bloch-Grüneisen fit. Although the Bloch-Grüneisen law apparently fits resistivity data up to ~220 K, it describes the coupling of charge carriers with the acoustic phonons only. Thus, for consistency, we limit the fitting of resistivity by eq. (1) to the low temperature $T^3$ range, characteristic of coupling of charge carriers with acoustic phonons, also seen in specific heat data. The best fit parameters are Debye temperature $\Theta_R \approx 200$ K and $\rho' \approx 1$ μΩ cm. The Debye temperatures $\Theta_D \approx 140\ K$ and $\Theta_R \approx 200\ K$ extracted from specific heat and resistivity, respectively, are in substantial agreement. With this $\rho'$ value, from eq. (14) we find $\lambda_{tr} \approx 0.11$. This value is close to the electron-phonon coupling constant ~0.21 calculated for 1111 iron pnictide compounds [26] and ~0.18 calculated for 122 compounds [27]. We point out that our experimental estimate refers to the coupling with acoustic phonons only, neglecting the contribution of optical phonons. From this $\lambda_{tr}$ value, the Bardeen-Cooper-Schrieffer evaluation of the critical temperature turns out negligible, suggesting that phonon coupling cannot be the main pairing mechanism into play. However, the relevance of the interaction with the phonons, emphasized by low impurity scattering, appears clearly from the analysis of the temperature dependence of normal state resistivity and Seebeck coefficient, placing this compound somewhat midway between conventional superconductors, such as A15 and MgB$_2$, and its peers iron-based unconventional superconductors. Indeed, even if it seems that in LaFe$_2$As$_2$ superconductivity is likely unconventional in nature, coupling with phonons could play a role, not only in the normal state, but also in the superconducting mechanisms, possibly accounting for the peculiar behavior of being superconducting far beyond the highly overdoped regime.

## 4. Conclusions

In this work, we present the first characterization of normal state transport and thermal properties of the LaFe$_2$As$_2$ superconducting compound, a highly metallic member of the 122 family of iron-based compounds. We combine specific heat, resistivity, magnetoresistivity, Hall effect, and Seebeck effect data to carry out a self-consistent data analysis in a two-band framework. We find evidence that although transport is dominated by an electron band, a hole band at the Fermi level does make a significant contribution to transport. Indeed, this compound is highly compensated, with electron and hole bands having equal carrier densities within ~1% and high effective masses ~3 m$_0$. Most remarkably, this compound is highly overdoped, with ~0.42 electrons per unit cell at 50 K. This finding challenges the usual belief that the superconducting dome in the phase diagram of 122 iron-based superconductors typically ends in correspondence of at 0.2 electrons per unit cell. As much remarkably, opposite to the behavior of its unconventional superconducting pnictide peers, normal state transport in this compound exhibits some distinctive features of the phonon-coupled traditional superconductors, namely a Bloch-Grüneisen type resistivity, a Ioffe-Regel saturation of resistivity at high temperatures, a phonon drag contribution to the Seebeck effect. We estimate an electron phonon coupling $\lambda_{tr} \approx 0.11$, too small to account for any role of phonons in the pairing mechanism. A*b initio* calculations of the electronic structure in literature [10,12] indicate that this compound is similar to other 122 superconducting pnictides. From these seemingly contrasting premises, the superconducting mechanisms in LaFe$_2$As$_2$ appear to be a puzzle and should be further investigated by theoretical and experimental approaches.

**Acknowledgements**
This work was supported by JSPS KAKENHI Grant Number JP16H6439.